\documentclass[conference,letterpaper]{IEEEtran}
\usepackage{threeparttable}
\usepackage{multirow}
\usepackage[caption=false]{subfig} % Handles subfigures
\usepackage{tabu} % Extended tabular
\usepackage{algorithm2e} % For algorithms
\usepackage{xcolor}
\usepackage{colortbl} % Coloring tables
\usepackage{tabularx} % Better tables
\usepackage{graphicx} % For including images
\usepackage{pifont} % Extra symbols
\usepackage{amsmath, amssymb, amsfonts}
\usepackage{algorithmic}
\usepackage{textcomp}
\usepackage{adjustbox}
\usepackage{bm}
\usepackage{microtype}
\usepackage{hyperref}
\usepackage{url}
\usepackage{booktabs}
\def\BibTeX{{\rm B\kern-.05em{\sc i\kern-.025em b}\kern-.08em T\kern-.1667em\lower.7ex\hbox{E}\kern-.125emX}}

\begin{document}
\title{Exploring Liquid Neural Networks on Loihi-2}

\makeatletter
\newcommand{\linebreakand}{%
  \end{@IEEEauthorhalign}
  \hfill\mbox{}\par
  \mbox{}\hfill\begin{@IEEEauthorhalign}
}
\makeatother

\author{%
\IEEEauthorblockN{Wiktoria Agata Pawlak}
\IEEEauthorblockA{%
\textit{Tilburg University}\\
Tilburg, The Netherlands\\
Email: w.a.pawlak@tilburguniversity.edu}
\and
\IEEEauthorblockN{Murat Isik}
\IEEEauthorblockA{%
\textit{Stanford University}\\
Stanford, United States\\
Email: mrtisik@stanford.edu}
\and
\IEEEauthorblockN{Dexter Le}
\IEEEauthorblockA{%
\textit{Drexel University}\\
Philadelphia, United States\\
Email: dql27@drexel.edu}
\linebreakand % Use your custom command for line breaking
\IEEEauthorblockN{Ismail Can Dikmen\IEEEauthorrefmark{1}}
\IEEEauthorblockA{%
\textit{TEMSA R\&D Center}\\
Adana, Türkiye\\
Email: can.dikmen@temsa.com}
Corresponding author}
\thanks{\IEEEauthorrefmark{1}Corresponding author: Ismail Can Dikmen, can.dikmen@temsa.com}

\maketitle

\begin{abstract}
This study investigates the realm of liquid neural networks (LNNs) and their deployment on neuromorphic hardware platforms. It provides an in-depth analysis of Liquid State Machines (LSMs) and explores the adaptation of LNN architectures to neuromorphic systems, highlighting the theoretical foundations and practical applications. We introduce a pioneering approach to image classification on the CIFAR-10 dataset by implementing Liquid Neural Networks (LNNs) on state-of-the-art neuromorphic hardware platforms. Our Loihi-2 ASIC-based architecture demonstrates exceptional performance, achieving a remarkable accuracy of 91.3\% while consuming only 213 microJoules per frame. These results underscore the substantial potential of LNNs for advancing neuromorphic computing and establish a new benchmark for the field in terms of both efficiency and accuracy.
\end{abstract}


\section{Introduction}
\label{sec:introduction}
The resurgence of neural network research in the 1980s and 1990s, catalyzed by advancements such as backpropagation and the introduction of recurrent and convolutional architectures \cite{xu-1992, huang-2009, dreyfus-1990, ciesinger-1988}, revitalized the field and enabled the training of more intricate models, overcoming early limitations like the perceptron's linear separability constraint. This resurgence of neural network research has led to various applications, from image recognition to natural language processing. It has also opened new research areas, such as developing deep learning models capable of learning hierarchical representations of data.

Those improvements led to the development of the Liquid Neural Network (LNN), an evolution of the Recurrent Neural Network (RNN). Its ability to adapt dynamically to structural changes is distinguished by its liquid state. Time-series data can be processed efficiently and flexibly using this mechanism, inputs can be preserved, and network behavior and structure can be dynamically adjusted using this mechanism. When LNNs are used to handle sequential data with complex temporal patterns, they can be reconfigured according to data and task demands, thus enabling continuous learning without labeling and handling sequential data with complex temporal patterns. During training, LNNs determine their topology, which remains the same at the execution stage \cite{xu-1992, huang-2009, dreyfus-1990, ciesinger-1988}.

LNN models were developed using new mathematical formulations and connectivity patterns to improve energy efficiency and causality in robots and other devices. Despite significant advancements in neural network (NN) technologies, a substantial gap persisted between the computational intelligence of biological brains and the capabilities of deep learning models. This gap is particularly evident in areas such as representation, learning capacity, understanding and interacting with the world while capturing causality, learning abstract actions for reasoning or planning, and the efficiency and flexibility of achieving goals \cite{chahine-2023, hasani-2021}. To bridge this gap, researchers explored the fundamental elements of the neuronal system, including neural circuits, neurons, and synapses. The resulting neural network structure, more advanced, memory-conserving, and flexible, was inspired by the biological nervous system, notably those found in the \textit{C. elegans nematode}, recognized for its complex behaviors with a minimal number of neurons \cite{lechner-2019}. It has led to the development of models aiming to mimic the propagation of information between neurons\cite{chahine-2023}.

Biological systems possess continuous neural dynamics that are described by differential equations, capturing the complex, time-dependent nature of neuronal activity, unlike traditional deep learning systems. The release mechanisms of synapses, derived from biology, extend beyond scalar weights, introducing nonlinearity into the system. This process mimics neurotransmitter binding to receptors, enriching computational models. Furthermore, neural circuits are distinguished by their recurrent connections, sparsity, and innate memory capabilities \cite{hasani-2021, huynh2022implementing, isik2023design, isik2024neurosec}.

Liquid Time-Constant Networks (LTCNs) offer nearly continuous mappings with a small number of computational units \cite{hasani-2018}. Hasani’s subsequent work in 2021 further developed and optimized the LTC network model. Time series forecasting tasks were performed more accurately, and its potential for embedded systems was demonstrated \cite{hasani-2021}. In 2023, the LTC-SE algorithm was introduced to enhance flexibility and code organization while maintaining compatibility with embedded systems \cite{bidollahkhani-2023}. With these improvements, LTCNs are now regarded as superior time series processing and forecasting tools.

\begin{figure*}[htbp]
    \centering
    \includegraphics[width=\textwidth]{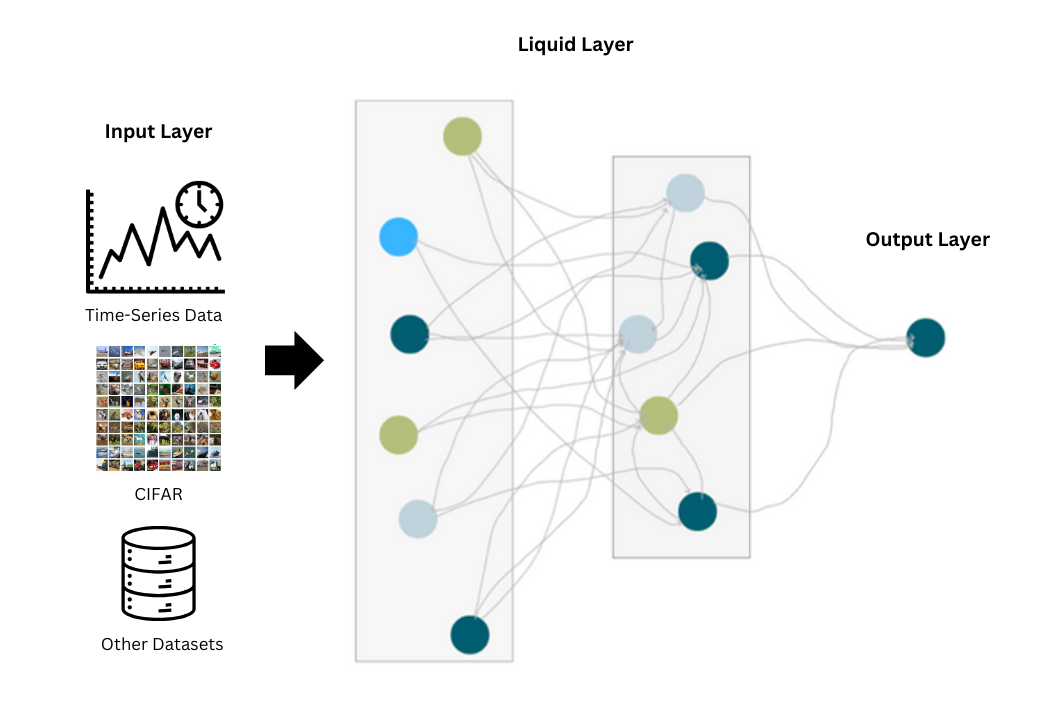}
    \caption{\textbf{Schematic Diagram of Liquid Neural Networks (LNNs). Figure illustrates simplified architecture of an LNN, starting with the input layer receiving time-series data or as in our case images from datasets like CIFAR-10. The data then flows into the liquid layer, where dynamic, non-linear processing occurs through a complex network of interconnected neurons. Finally, the processed information is relayed to the output layer.}}
    \label{fig1}
\end{figure*}

The goal of this research is to boost computational efficiency and adaptability by integrating LTCNs with neuromorphic hardware. In this paper, we examine LTCN compatibility and optimization within neuromorphic frameworks, address integration challenges, and quantify speed, energy efficiency, and intelligence simulation improvements using empirical analysis and case studies. In subsequent sections, we examine theoretical foundations, practical applications, challenges, and opportunities for adapting LTCN to neuromorphic hardware. Our goal is to raise awareness about LTCNs through a creative case study and practical application. We aim to illustrate the substantial contributions LTCNs have made to neuroscience and the practical economic value they represent.

\section{Understanding Liquid Neural Networks and Its Implementation}
\subsection{Theoretical Background}
Liquid Time Constant Neural Networks (LNNs) depart from conventional neural network architectures by drawing inspiration from biological systems. This shift has led to the development of continuous-time neural networks, which are distinguished by several key characteristics:

\subsubsection{Inspiration from biological models}
LNNs draw inspiration from Louis Lapicque's leaky integrator model and the Hodgkin-Huxley conductance-based synapse model, incorporating differential equations into activation functions for dynamic modeling of neuronal communication. It attempts to capture the neuronal membrane dynamics—the core difference from traditional NNs—by allowing the introduction of nonlinearities \cite{lapicque:jppg1907, hodgkin-1952, hasani-2020}. Unlike conventional NNs, where nonlinearity is typically localized within the neurons, LNNs distribute this nonlinearity across the network, especially in synaptic interactions. This shift leads to more flexible representations of synaptic interactions, similar to those exhibited by biophysical ion channels as described by the Hodgkin-Huxley model, and offers a computational advantage in handling temporal dynamics. Consequently, LNNs process temporal sequences with higher accuracy and efficiency compared to traditional RNNs, which often struggle with accurately capturing the temporal dependencies due to their fixed architectures \cite{saha-2023, hasani-2020,lechner-2019}.

\subsubsection{Dynamic system properties}
Building on their architectural principles, LNNs exhibit dynamic system properties grounded in ordinary differential equations (ODEs). The neural network function, f, and the system’s state, x(t), include nonlinearities in their coefficients, contributing to the network's fluidity and adaptivity. In LNNs, time constants are coupled with differential variables. This facilitates system coherence over time, different from traditional RNNs based on recurrent connections and complex architectures \cite{nielsen-2023, niu-2019}.

\subsubsection{Stability without recurrency}
Besides their behavior and state stability, LNNs have a time constant that governs them. The system maintains coherence over time, adapting dynamically to incoming data by avoiding recurrent connections. In contrast, traditional RNNs, with their traditional architecture and mechanisms, often cannot achieve the same level of stability. LNNs use a mathematical formulation that requires less computation and stabilizes neurons while training. Unlike typical neural networks, they are capable of adapting to changing situations after training by using differential equations that can be dynamically adjusted \cite{hasani-2020, nielsen-2023}.

\subsubsection{The Continuous-time Neural Network Framework}
LNNs characterized by defined layers, widths, and activation functions can be used to parameterize derivatives of the hidden state for continuous-time processes. This configuration does not constrain traditional models like residual networks, which utilize discrete time steps. Furthermore, this model provides flexibility not limited by fixed time steps, allowing networks to function over variable depths and potentially reach infinite depths within a single processing cycle. As a result of this flexibility, continuous-time networks are particularly appropriate for modeling sequential behavior with natural fluidity compared with their discrete-time counterparts \cite{hasani-2020, chahine-2023}.

\subsubsection{Efficiency and memory conservation}
LNNs work by dynamically adjusting synaptic weights and underlying equations to respond to new or noisy inputs, which is memory-saving and computationally efficient. Traditional RNNs are limited by fixed synaptic weights, limiting their ability to adapt, emphasizing LNNs' superior modeling of biological learning and memory. The memory-conserving design reflects the selective strengthening or weakening of synapses in biological neural networks to inform LNNs' selective information retention approach based on synaptic plasticity. By preserving critical data from past inputs, LNNs can efficiently discard irrelevant or redundant information while retaining critical information from past inputs. A combination of dynamic gating mechanisms and adaptive thresholding ensures that only salient information influences the network's state over time, ensuring that performance is maintained across tasks, even those requiring long-term temporal dependencies \cite{hasani-2018, hasani-2020, lechner-2019}.

\subsection{Practical Implementation}
Most LNNs have been deployed through software simulation, while emerging hardware adaptations demonstrate their adaptability, compact size, and ability to process time-series data. LNNs are not only capable of learning, but they are also robust, flexible, and interpretable because of their built-in variability and ability to adapt their equations based on the input they observe, making them well suited for analyzing time-varying phenomena such as electric power grids, financial transactions, and weather forecasting.

LNNs have proven useful in robotics, particularly in safety-critical systems, where traditional networks fail because of their task-oriented rather than context-oriented approach. Researchers at MIT have studied robotic image processing tasks such as object tracking, segmentation, and recognition. Their work on autonomous systems has demonstrated the ability of LNNs to cope with unpredictable environmental interactions, reducing processing errors \cite{hasani-2020, lechner-2019}. In commercial applications, LNNs are exemplified by Liquid.ai, a company that focuses on workflow management efficiency through LNNs.

A compelling example of LNN application for autonomous vehicles is the CALNet project. By integrating 3D LiDAR and 2D cameras and leveraging the hybrid spatial pyramid pooling (HSPP) method along with liquid time constant networks (LTC), CALNet has demonstrated superior efficiency in automating sensor calibration tasks within dynamic environments compared to traditional deep learning methods. This advancement is critical for achieving accurate multi-sensor data fusion, a cornerstone of sophisticated autonomous systems \cite{shang-2022}.

Using LTC networks, multiple-input multiple-output (MIMO) wireless communication systems can overcome the hurdles presented by legacy Channel State Information (CSI). Compared to traditional RNN methodologies, LTC-based techniques have superior accuracy, stability, and adaptability, outperforming existing algorithms by 10-40\%. This capability to navigate the dynamic landscape of wireless communication environments underscores the potential of LTC networks in this domain \cite{yin-2021}.

Demonstrating adaptability and efficiency, LTC networks excel at predicting congestion and duration on millimeter wave (mmWave) communication links without the need for retraining or scenario-specific data. Moreover, these networks contribute to advancing low-latency, high-reliability communication systems by mitigating reliability and signal congestion challenges inherent to mmWave environments \cite{nielsen-2023}.

Using their innovative LTC-SE neural network architecture, Bidollahkhani et al. have proven to be more accurate in time series prediction, human behavior recognition, and other domains than their predecessor, the LTC. By optimizing for resource-constrained environments and incorporating elements from diverse neural network architectures, LTC-SE demonstrates superior computational efficiency, accuracy, and memory utilization, making it a perfect candidate for deployment on platforms such as neuromorphic chips \cite{bidollahkhani-2023}.

The healthcare domain has witnessed a transformative application of LNNs through Closed-Form Continuous-Time Liquid Neural Networks (CfCs). These networks, in conjunction with knowledge graphs, underpin a comprehensive framework for analyzing complex patient data. This innovation empowers real-time analytics, enabling earlier diagnoses, tailored treatment plans, and optimized surgical interventions. The computational efficiency and adaptability of CfCs have been instrumental in constructing dynamic patient health models, marking a pivotal advancement in biomedical applications of LNNs \cite{nye-2023}.

\begin{figure*}[htbp]
    \centering
    \includegraphics[width=0.8\textwidth]{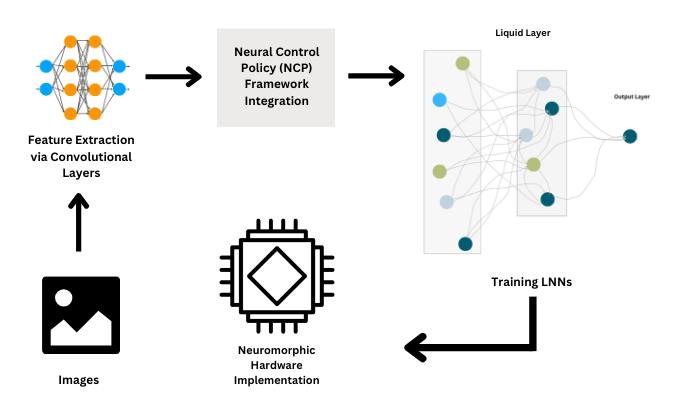}
    \caption{\textbf{LNNs Hardware Implementation. Figure illustrates the data processing flow of implementing an LNN into neuromorphic hardware, beginning with the CIFAR-10 dataset at the bottom which we deployed, which undergoes feature extraction via convolutional layers. The extracted features are then integrated within the Neural Circuit Policy (NCP) Framework, a decision-making system that steers the data through the liquid layer illustrated in the top-right. After training, the LNN is implemented on neuromorphic hardware, symbolized by the chip icon.}}
    \label{fig2}
\end{figure*}

\subsection{Adapting LNNs for Neuromorphic Systems}
Despite the recent advancements in LNNs, many research gaps and unexplored areas could significantly benefit various fields and applications, particularly regarding their integration with neuromorphic hardware. LNNs use 1-3 orders of magnitude fewer parameters \cite{hasani-2020, hasani-2021}, enabling the algorithm to run on smaller devices, thus reducing power consumption and computational cost. This efficiency makes their adaptation to neuromorphic computing much more impactful. However, a key challenge is ensuring compatibility between the biologically inspired mathematical equations of LNNs and the specific features of neuromorphic architectures to optimize computational capabilities and achieve energy-efficient operations.

Addressing these challenges requires strategies like hardware customization, developing new types of chips tailored to support the algorithm and allow it to scale up for faster computation of more data. Encoding and decoding techniques that are efficient, as well as algorithm-hardware co-design, are also essential. In order to adapt LNN algorithms successfully to neuromorphic systems, it is essential to understand both the dynamic properties of LNN algorithms and the physical constraints and capabilities of neuromorphic chips designed with current technology.

It is necessary to reconceive traditional computing models in order to create chips that mimic both information processing and neuronal behavior. This approach, similar to how the algorithm in LNNs was mapped based on \textit{C.elegans }neurons, takes into account not only computational processes but also the whole dynamics of neuronal interaction \cite{lechner-2019}. LNNs' adaptability, parallelism, and low-power operation will be fully exploited by neuromorphic hardware. Therefore, LNNs may be capable of rivaling, and even exceeding, the performance of current algorithms such as Spiking Neural Networks (SNNs).

\subsection{Challenges and Solutions}
Neuromorphic hardware requires LNNs that are compatible, efficient, and scalable. Research efforts are currently focused on developing learning algorithms and optimizing neural structures and hyperparameters. Enhanced accuracy, scalability, and computational efficiency will enable LNNs to be more widely used. Due to their architecture, LNNs offer significant advantages, but their integration with neuromorphic hardware, which usually uses SNNs, also presents technical and theoretical challenges.

\begin{itemize}
    \item Model complexity and scalability: 
LNNs are known for processing input both dynamically and temporally. Due to this, their internal states are constantly changing, adding to their complexity. It is difficult to scale these networks using neuromorphic technology designed for more straightforward, static neural networks. Liquid state dynamics must be preserved despite hardware limitations with constrained memory and processing power \cite{hasani-2020}.

    \item Temporal dynamics representation: 
It is crucial for LNNs to represent and process time accurately. The implementation of these dynamics on neuromorphic hardware, which operates discretely, poses significant challenges. The continuous-time operations of LNNs must be effectively translated into discrete time steps without losing essential temporal information \cite{lechner-2019}.

    \item Energy efficiency: 
Neuromorphic computing's appeal includes its potential for high computational efficiency at low energy costs. However, the dynamic and complex nature of LNNs may increase power consumption due to continuous information processing and the need to maintain the liquid state. Optimizing energy efficiency while preserving LNNs' computational capabilities, for example, through techniques like compression, is critical \cite{hasani-2021}.

The scalability, adaptability, and resource efficiency challenges have already been addressed in some research efforts, particularly on resource-constrained embedded systems with the LTC-SE neural network algorithm, an enhanced version of the Liquid Time Constant neural network algorithm\cite{bidollahkhani-2023}.
\end{itemize}

\subsection{Integration Challenges}
Implementing LNNs on neuromorphic hardware presents specific challenges related to compatibility, efficiency, and scalability. One such compatibility challenge is the computational complexity of LNNs which can impact performance \cite{hasani-2021}. Subsequently, efficiency is a challenge in the energy domain where the overall Energy Delay Product (EDP) should be reduced to the minimum. Additionally, latency is important to enhance efficiency. One study observed Intel's Loihi neuromorphic computing chip which distinguished CPU and GPU performance increase on closed-loop sensing through SNNs \cite{daves2021computingloihi}. Despite differences in architectural designs, LNNs face similar computational complexity challenges that complement the imperative need for lower latency. These challenges impact the performance and practicality of LNNs in real-world applications, necessitating innovative solutions to integrate these networks with neuromorphic systems. It is essential to address these problems in order to make better use of LNNs in neuromorphic computing, especially in cases where real-time processing and energy efficiency are critical.\cite{wu2023perfecting}.

\subsection{Case Studies and Applications}
There are several case studies and practical applications where algorithms similar to LNNs have been successfully implemented on neuromorphic hardware. For Intel's Loihi, one study employs the Liquid State Machine (LSM) simulated on MATLAB to measure the memory metric proposed by Gorad et al. on Loihi. The memory metric observed validated the feasibility of designing LSMs for large-scale networks \cite{patel-2022}. Additionally, another study employs an extended LSM (ELSM) which increases accuracy in speech recognition. ELSM aims to provide additional benefits on top of the base LSM, such as balancing excitatory and inhibitory presynaptic currents to maximize neural efficiency \cite{nihExtendedLiquid}. These examples demonstrate the practical benefits and breakthroughs highlighting algorithms' potential in fields such as cognitive computing, speech recognition, and real-time data analysis. However, these implementations also reveal limitations and areas for improvement, guiding future research in optimizing them for neuromorphic systems.

\subsection{Optimization Opportunities}
Given the adaptive nature of both LNNs and neuromorphic hardware, integrating both components can offer great potential in improving current practices. There are studies that show that LSMs cannot maintain robustness if damages are applied to the neurons \cite{hazan-2012}. Robustness is a great optimization challenge as failure to abide a healthy state can impact performance and accuracy. Another study scrutinizes the proposition of the "stochastic bit" which has been observed to improve energy efficiency through a spiking neural network utilizing on-chip learning \cite{koo-2020}. A variable probability determines the degree of toggling between low and high states in a stochastic bit. The stochastic bit, realized through the integration of PMOS header and NMOS footer transistors, enables a mapping of input sNeurons, representing image data, to output sNeurons via binary weights. This stochastic mechanism can trigger the firing of output sNeurons, with the computational efficiency of weighted inputs benefiting from AND operations. These architectural elements present substantial opportunities to elevate neuromorphic technology beyond contemporary benchmarks.

\section{Method}
We propose the LNN module, a novel architecture designed to extract and analyze spatial features from datasets like photographs. By cascading convolutional layers to progressively reduce data dimensionality, the model generates abstract representations that inform decision-making through the Neural Circuit Policy (NCP). A cyclic training regimen, employing backpropagation and Adam optimization, refines model parameters to enhance predictive accuracy. To address computational constraints in real-world deployments, particularly on edge devices, the model undergoes quantization and transformation processes. Quantization selectively reduces parameter sensitivity, minimizing memory footprint and computational overhead, while transformation adapts the model to diverse hardware platforms. This integrated approach positions the LNN module as a promising solution for spatial data analysis in resource-limited environments.

\begin{figure*}[htbp]
    \centering
    \includegraphics[width=0.8\textwidth]{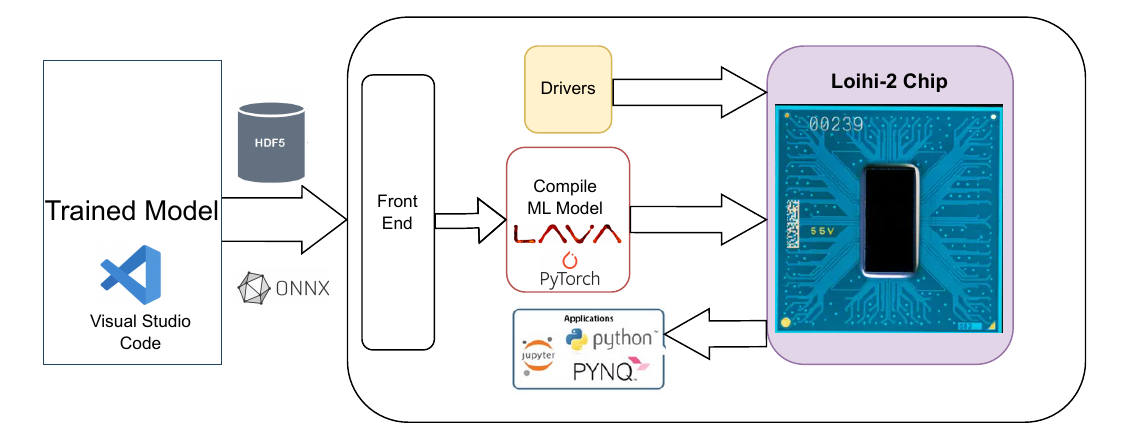}
    \caption{\textbf{Block Diagram of Implementation.}}
    \label{fig3}
\end{figure*}

The diagram in \autoref{fig3} outlines implementing a machine learning model on the Loihi-2 chip, specialized in neuromorphic computing. This process starts with the trained model in a development environment and proceeds through several readiness and compatibility checks using a front-end user interface. Then, using certain drivers and tools, the model is compiled into an executable form for the Loihi-2 processor under LAVA Framework. Lastly, the model is running on neuromorphic hardware and prepared to carry out tasks with biological neural system-like efficiency. It makes use of the special powers of neuromorphic computing to process data in a way that is inspired by the human brain.

We utilized Python to execute implementations on the GPU. The study leveraged the computational prowess of NVIDIA's GeForce RTX 3060 GPU which is optimized for different tasks, ensuring an efficient execution of our implementations for each neural network model.

\subsection{Hardware Accelerator MAC Components}
\begin{enumerate}
    \item \textbf{LNN Embedding:}
    The embedding in LNNs transforms input data into dense vector embeddings. For input features of dimension \(D\) and embeddings of dimension \(S\), the MAC operations are:
    \begin{equation}
        \text{MAC}_{\text{Embedding}} = D \times S
    \end{equation}

    \item \textbf{Dynamic Adaptation Layer:}
    This layer adapts the network's architecture based on the input data over time \(t\), with \(A\) representing the adaptation cost:
    \begin{equation}
        \text{MAC}_{\text{Adaptation}} = A \times t
    \end{equation}

    \item \textbf{LNN Processing Layer:}
    For processing layers, where \(N\) is the number of active neurons, \(C\) is the average number of connections per neuron, and \(t\) represents time:
    \begin{equation}
        \text{MAC}_{\text{Processing}} = N \times C \times t
    \end{equation}
\end{enumerate}

The total MAC operations for an LNN is the sum of the operations for each component:
\begin{equation}
    \text{MAC}_{\text{Total}} = \text{MAC}_{\text{Embedding}} + \text{MAC}_{\text{Adaptation}} \\
    + \text{MAC}_{\text{Processing}}
\end{equation}

\subsection{Throughput Analysis}
The MAC counts, as derived from neural network libraries, and the latency or simulation time, primarily influenced by the hardware components, can be used to compute the throughput of the system:

\begin{equation}
\text{Latency} = \frac{\text{Total Inference Time}}{\text{Total number of inference samples}}
\label{eq:sim}
\end{equation}

This relationship can be represented as:
\begin{equation}
    \text{Throughput} = \frac{\# \text{MACs}}{\text{Latency}}
    \label{eq:throughput}
\end{equation}

\begin{table}[ht]
    \caption{Comparative Analysis.}
    \renewcommand{\arraystretch}{1.5}
    \setlength{\tabcolsep}{5pt}
    \centering
    \begin{threeparttable}
    {\fontsize{9}{10}\selectfont
        \begin{tabular}{c|cccc}
            \hline
            Model Characteristics & DNN & CNN & SNN & \textbf{LNN} \\
            \hline
            
            Accuracy (\%) & 85.1 & 89.0 & 82.5 & \textbf{91.3} \\
            MAC (GOP) & 1.32 & 1.05 & 0.95 & \textbf{0.85} \\
            Latency (ms) & 24.2 & 19.5 & 35.0 & \textbf{15.2} \\
            \hline
            *Power Efficiency & 10.4 & 15.0 & 8.2 & \textbf{25.3} \\
            \hline
        \end{tabular}}
        \begin{tablenotes}
            \item *Power Efficiency unit is GOP/s/W.
            \item *CIFAR-10 dataset was implemented on RTX 3060.
        \end{tablenotes}
    \end{threeparttable}
    \label{tab:comparison_others}
\end{table}

\section{Results}
\autoref{tab:comparison_others} delineates a comprehensive comparison between the proposed LNN framework and other neural network models, including DNN, CNN, and SNN. This analysis underscores the advantages and design paradigms tailored to their respective application domains with the LNN model demonstrating exceptional proficiency across various performance metrics. Notably, the LNN framework achieves accuracy of 91.3\%, surpassing its counterparts. This improved performance validates the LNN design and demonstrates that its robustness and durability make it suitable for demanding pattern recognition applications. Furthermore, among the evaluated models, the LNN model exhibits the lowest latency of 15.2 ms, demonstrating its exceptional efficiency and real-time data processing potential. Applications that need to make quick decisions based on real-time data streams, such as real-time monitoring and autonomous driving systems, will particularly benefit from this functionality. The optimal architecture of the LNN is revealed by the computational complexity measured in MAC operations (0.85 GOP). This design maintains high throughput while balancing the processing power and computational demand. This optimization is important for deployment in contexts with constrained computational resources. Furthermore, a key feature for long-term operation in power-sensitive applications is the large power efficiency of the LNN model (25.3 GOP/s/W). Due to its efficiency, the LNN framework can be integrated with IoT sensors, mobile devices, and other edge computing devices, allowing advanced computational capabilities with low energy usage. The LNN framework demonstrates improved accuracy, efficiency, and computational optimization. It's performance, efficiency, and power consumption make it the best choice for many applications, from embedded devices to large data centers.

\begin{table}[ht]
    \caption{Specifications of the Neuromorphic Hardware Device Utilized in the Study}
    \renewcommand{\arraystretch}{1.5}
    \setlength{\tabcolsep}{5pt}
    \centering
    \begin{threeparttable}
    {\fontsize{9}{10}\selectfont
        \begin{tabular}{l|c}
            \hline
            \textbf{Specification} & \textbf{Loihi 2 (Intel)} \\
            \hline
            Technology Node & 7nm \\
            Core Count & 128 \\
            Precision & Fixed 32 \\
            Operation Frequency & Variable \\
            Memory Technology & On-chip \\
            Power Consumption & Low Power \\
            Release Year & 2022 \\
            \hline
        \end{tabular}}
    \end{threeparttable}
    \label{table:neuromorphic_hardware_specs}
\end{table}

\begin{table}[ht]
    \renewcommand{\arraystretch}{1.8}
    \setlength{\tabcolsep}{5pt}
    \centering
    \caption{Comparisons with state of art implementations for CIFAR-10 dataset.}
    \label{table:table_3}
    \resizebox{\linewidth}{!}{
    \begin{tabular}{c|c|c|c|c}
    \hline
    \textbf{Work} & \textbf{Hardware} & \textbf{Technology (nm)} & \textbf{Accuracy (\%)} & \textbf{Efficiency (J/frame)} \\ \hline
    \cite{zhao2017accelerating} & FPGA & 28 & 88 & $27.9m$  \\ \hline
    \cite{esser2016cover} & ASIC & 28 & 83 & $164\mu$ \\ \hline
    \cite{bankman2018always} & ASIC & 28 & 86 & $3.8\mu$  \\ \hline
    \cite{moons2018binareye} & ASIC & 28 & 86 & $14.4\mu$  \\ \hline
    \cite{hsu2023fully} & FPGA & 16 & 88 & $13.7\mu$  \\ \hline
    \textbf{Our work} & \textbf{ASIC (Loihi-2)} & \textbf{7} & \textbf{91.3} & \textbf{213$\mu$} \\ \hline
    \end{tabular}
    }
\end{table}

\section{Conclusion}
This manuscript has proven the potential of LNNs in the context of neuromorphic hardware integration. It has explored the transformative potential of LNNs in reshaping computational paradigms to more closely mimic the operational efficiency and cognitive capabilities of the human brain, enhancing computational intelligence and reducing energy consumption. Furthermore, this study has projected the future trajectory of LNN research within the computational sphere. Our evaluations conducted on various neuromorphic chips utilizing the CIFAR-10 dataset have yielded promising outcomes, particularly with advanced semiconductor technologies achieving a commendable accuracy rate of 91.3\% in image classification tasks alongside minimal energy expenditure. These findings underscore the viability and progressive impact of LNNs on neuromorphic computing, setting a precedent for future advancements. Looking forward, the horizon of LNN research is vast and ripe with opportunities. The preliminary success achieved in this and related studies serves as a foundation upon which further explorations into the architectural, algorithmic, and application-specific optimizations of LNNs should be built. There exists a palpable potential to extend the application of LNNs beyond mere image classification, venturing into domains such as robotics, autonomous systems, and real-time analytics, which could significantly benefit from the adaptability and efficiency of LNN-based models. It articulates the premise that LNNs hold the key to advancing computing systems towards achieving brain-like efficiency and intelligence. While numerous challenges remain in realizing the full spectrum of LNN capabilities, the pathway towards integrating these advanced neural network models into practical, energy-efficient computing architectures appears both viable and promising. The continued exploration and development of LNNs are anticipated to spearhead significant innovations in artificial intelligence, paving the way for more sustainable and intelligent computing solutions.

\bibliographystyle{IEEEtran}
\bibliography{external}

\begin{thebibliography}{10}
\providecommand{\url}[1]{#1}
\csname url@samestyle\endcsname
\providecommand{\newblock}{\relax}
\providecommand{\bibinfo}[2]{#2}
\providecommand{\BIBentrySTDinterwordspacing}{\spaceskip=0pt\relax}
\providecommand{\BIBentryALTinterwordstretchfactor}{4}
\providecommand{\BIBentryALTinterwordspacing}{\spaceskip=\fontdimen2\font plus
\BIBentryALTinterwordstretchfactor\fontdimen3\font minus \fontdimen4\font\relax}
\providecommand{\BIBforeignlanguage}[2]{{%
\expandafter\ifx\csname l@#1\endcsname\relax
\typeout{** WARNING: IEEEtran.bst: No hyphenation pattern has been}%
\typeout{** loaded for the language `#1'. Using the pattern for}%
\typeout{** the default language instead.}%
\else
\language=\csname l@#1\endcsname
\fi
#2}}
\providecommand{\BIBdecl}{\relax}
\BIBdecl

\bibitem{xu-1992}
\BIBentryALTinterwordspacing
L.~Xu, S.~Klasa, and A.~Yuille, ``{RECENT ADVANCES ON TECHNIQUES OF STATIC FEEDFORWARD NETWORKS WITH SUPERVISED LEARNING},'' \emph{International Journal of Neural Systems}, vol.~03, no.~03, pp. 253--290, 1 1992. [Online]. Available: \url{https://doi.org/10.1142/s0129065792000218}
\BIBentrySTDinterwordspacing

\bibitem{huang-2009}
\BIBentryALTinterwordspacing
Y.~Huang, ``{Advances in artificial Neural Networks – Methodological development and application},'' \emph{Algorithms}, vol.~2, no.~3, pp. 973--1007, 8 2009. [Online]. Available: \url{https://doi.org/10.3390/algor2030973}
\BIBentrySTDinterwordspacing

\bibitem{dreyfus-1990}
\BIBentryALTinterwordspacing
S.~E. Dreyfus, ``{Artificial neural networks, back propagation, and the Kelley-Bryson gradient procedure},'' \emph{Journal of Guidance Control and Dynamics}, vol.~13, no.~5, pp. 926--928, 9 1990. [Online]. Available: \url{https://doi.org/10.2514/3.25422}
\BIBentrySTDinterwordspacing

\bibitem{ciesinger-1988}
\BIBentryALTinterwordspacing
J.~Ciesinger, ``{Neural nets},'' \emph{SIGACT news}, vol.~19, no. 3-4, pp. 46--47, 11 1988. [Online]. Available: \url{https://doi.org/10.1145/58395.58398}
\BIBentrySTDinterwordspacing

\bibitem{chahine-2023}
\BIBentryALTinterwordspacing
M.~Chahine, R.~Hasani, P.~S.~S. Kao, A.~Ray, R.~Shubert, M.~Lechner, A.~Amini, and D.~Rus, ``{Robust flight navigation out of distribution with liquid neural networks},'' \emph{Science robotics}, vol.~8, no.~77, 4 2023. [Online]. Available: \url{https://doi.org/10.1126/scirobotics.adc8892}
\BIBentrySTDinterwordspacing

\bibitem{hasani-2021}
\BIBentryALTinterwordspacing
R.~Hasani, M.~Lechner, A.~Amini, D.~Rus, and R.~Grosu, ``{Liquid time-constant networks},'' \emph{Proceedings of the ... AAAI Conference on Artificial Intelligence}, vol.~35, no.~9, pp. 7657--7666, 5 2021. [Online]. Available: \url{https://doi.org/10.1609/aaai.v35i9.16936}
\BIBentrySTDinterwordspacing

\bibitem{lechner-2019}
\BIBentryALTinterwordspacing
M.~Lechner, R.~Hasani, M.~Zimmer, T.~A. Henzinger, and R.~Grosu, ``{Designing Worm-inspired Neural Networks for Interpretable Robotic Control},'' \emph{IEEE}, 5 2019. [Online]. Available: \url{https://doi.org/10.1109/icra.2019.8793840}
\BIBentrySTDinterwordspacing

\bibitem{huynh2022implementing}
P.~K. Huynh, M.~L. Varshika, A.~Paul, M.~Isik, A.~Balaji, and A.~Das, ``Implementing spiking neural networks on neuromorphic architectures: A review,'' \emph{arXiv preprint arXiv:2202.08897}, 2022.

\bibitem{isik2023design}
M.~Isik, K.~Inadagbo, and H.~Aktas, ``Design optimization for high-performance computing using fpga,'' in \emph{Annual International Conference on Information Management and Big Data}.\hskip 1em plus 0.5em minus 0.4em\relax Springer, 2023, pp. 142--156.

\bibitem{isik2024neurosec}
M.~Isik, H.~Vishwamith, Y.~Sur, K.~Inadagbo, and I.~C. Dikmen, ``Neurosec: Fpga-based neuromorphic audio security,'' in \emph{International Symposium on Applied Reconfigurable Computing}.\hskip 1em plus 0.5em minus 0.4em\relax Springer, 2024, pp. 134--147.

\bibitem{hasani-2018}
\BIBentryALTinterwordspacing
R.~M. Hasani, M.~Lechner, A.~Amini, D.~Rus, and R.~Grosu, ``{Liquid time-constant recurrent neural networks as universal approximators},'' 11 2018. [Online]. Available: \url{https://arxiv.org/abs/1811.00321v1}
\BIBentrySTDinterwordspacing

\bibitem{bidollahkhani-2023}
\BIBentryALTinterwordspacing
M.~Bidollahkhani, F.~Atasoy, and H.~Abdellatef, ``{LTC-SE: Expanding the potential of Liquid Time-Constant Neural Networks for scalable AI and embedded systems},'' \emph{arXiv (Cornell University)}, 4 2023. [Online]. Available: \url{https://arxiv.org/abs/2304.08691}
\BIBentrySTDinterwordspacing

\bibitem{lapicque:jppg1907}
L.~Lapicque, ``Recherches quantitatives {sur} {l'}excitation electrique des nerfs trait{\'e}e comme une polarization,'' \emph{Journal de Physiologie et Pathologie General}, vol.~9, pp. 620--635, 1907.

\bibitem{hodgkin-1952}
\BIBentryALTinterwordspacing
A.~L. Hodgkin and A.~Huxley, ``{A quantitative description of membrane current and its application to conduction and excitation in nerve},'' \emph{The Journal of Physiology}, vol. 117, no.~4, pp. 500--544, 8 1952. [Online]. Available: \url{https://doi.org/10.1113/jphysiol.1952.sp004764}
\BIBentrySTDinterwordspacing

\bibitem{hasani-2020}
\BIBentryALTinterwordspacing
R.~Hasani, M.~Lechner, A.~Amini, D.~Rus, and R.~Grosu, ``{Liquid time-constant networks},'' \emph{arXiv (Cornell University)}, 6 2020. [Online]. Available: \url{https://arxiv.org/abs/2006.04439}
\BIBentrySTDinterwordspacing

\bibitem{saha-2023}
\BIBentryALTinterwordspacing
N.~Saha, A.~Swetapadma, and M.~Mondal, ``{A Brief Review on Artificial Neural Network: Network Structures and Applications},'' \emph{IEEE}, 3 2023. [Online]. Available: \url{https://doi.org/10.1109/icaccs57279.2023.10112753}
\BIBentrySTDinterwordspacing

\bibitem{nielsen-2023}
\BIBentryALTinterwordspacing
M.~H. Nielsen, C.-Y. Yeh, M.~Shen, and M.~Médard, ``{Blockage prediction in directional MMWAve links using liquid time constant network},'' \emph{IEEE}, 9 2023. [Online]. Available: \url{https://doi.org/10.1109/irmmw-thz57677.2023.10299092}
\BIBentrySTDinterwordspacing

\bibitem{niu-2019}
\BIBentryALTinterwordspacing
M.~Y. Niu, L.~Horesh, and I.~Chuang, ``{Recurrent neural networks in the eye of differential equations},'' 4 2019. [Online]. Available: \url{https://arxiv.org/abs/1904.12933}
\BIBentrySTDinterwordspacing

\bibitem{shang-2022}
\BIBentryALTinterwordspacing
H.~Shang and B.~Hu, ``{CALNET: LiDAR-Camera Online calibration with channel attention and Liquid Time-Constant Network},'' \emph{2022 26th International Conference on Pattern Recognition (ICPR)}, 8 2022. [Online]. Available: \url{https://doi.org/10.1109/icpr56361.2022.9956145}
\BIBentrySTDinterwordspacing

\bibitem{yin-2021}
\BIBentryALTinterwordspacing
H.~Yin, Y.~Zhou, L.~Cao, and Y.~Xu, ``{Channel Prediction with Liquid Time-Constant Networks: An Online and Adaptive Approach},'' \emph{2021 IEEE 94th Vehicular Technology Conference (VTC2021-Fall)}, 9 2021. [Online]. Available: \url{https://doi.org/10.1109/vtc2021-fall52928.2021.9625323}
\BIBentrySTDinterwordspacing

\bibitem{nye-2023}
\BIBentryALTinterwordspacing
L.~Nye, ``{Digital twins for patient care via knowledge graphs and Closed-Form Continuous-Time liquid neural networks},'' \emph{arXiv (Cornell University)}, 7 2023. [Online]. Available: \url{https://arxiv.org/abs/2307.04772}
\BIBentrySTDinterwordspacing

\bibitem{daves2021computingloihi}
M.~Davies, A.~Wild, G.~Orchard, Y.~Sandamirskaya, G.~A.~F. Guerra, P.~Joshi, P.~Plank, and S.~R. Risbud, ``Advancing neuromorphic computing with loihi: A survey of results and outlook,'' \emph{Proceedings of the IEEE}, vol. 109, no.~5, pp. 911--934, 2021.

\bibitem{wu2023perfecting}
J.~Wu and M.~Gu, ``Perfecting liquid-state theories with machine intelligence,'' \emph{The Journal of Physical Chemistry Letters}, vol.~14, pp. 10\,545--10\,552, 2023.

\bibitem{patel-2022}
\BIBentryALTinterwordspacing
R.~Patel, V.~Saraswat, and U.~Ganguly, \emph{{Liquid State Machine on LOIHI: Memory Metric for Performance Prediction}}.\hskip 1em plus 0.5em minus 0.4em\relax Springer Nature Switzerland, 1 2022. [Online]. Available: \url{https://doi.org/10.1007/978-3-031-15934-3_57}
\BIBentrySTDinterwordspacing

\bibitem{nihExtendedLiquid}
L.~Deckers, I.~J. Tsang, W.~Van~Leekwijck, and S.~Latr{\'e}, ``Extended liquid state machines for speech recognition,'' \emph{Frontiers in Neuroscience}, vol.~16, p. 1023470, 2022.

\bibitem{hazan-2012}
\BIBentryALTinterwordspacing
H.~Hazan and L.~M. Manevitz, ``{Topological constraints and robustness in liquid state machines},'' \emph{Expert systems with applications}, vol.~39, no.~2, pp. 1597--1606, 2 2012. [Online]. Available: \url{https://doi.org/10.1016/j.eswa.2011.06.052}
\BIBentrySTDinterwordspacing

\bibitem{koo-2020}
\BIBentryALTinterwordspacing
M.~Koo, G.~Srinivasan, Y.~Shim, and K.~Roy, ``{SBSNN: Stochastic-BITS enabled binary spiking neural network with On-Chip learning for energy efficient neuromorphic computing at the edge},'' \emph{IEEE transactions on circuits and systems. I, Regular papers (Print)}, vol.~67, no.~8, pp. 2546--2555, 8 2020. [Online]. Available: \url{https://doi.org/10.1109/tcsi.2020.2979826}
\BIBentrySTDinterwordspacing

\bibitem{zhao2017accelerating}
R.~Zhao, W.~Song, W.~Zhang, T.~Xing, J.-H. Lin, M.~Srivastava, R.~Gupta, and Z.~Zhang, ``Accelerating binarized convolutional neural networks with software-programmable fpgas,'' in \emph{Proceedings of the 2017 ACM/SIGDA International Symposium on Field-Programmable Gate Arrays}, 2017, pp. 15--24.

\bibitem{esser2016cover}
S.~K. Esser, P.~A. Merolla, J.~V. Arthur, A.~S. Cassidy, R.~Appuswamy, A.~Andreopoulos, D.~J. Berg, J.~L. McKinstry, T.~Melano, D.~R. Barch \emph{et~al.}, ``From the cover: Convolutional networks for fast, energy-efficient neuromorphic computing,'' \emph{Proceedings of the National Academy of Sciences of the United States of America}, vol. 113, no.~41, p. 11441, 2016.

\bibitem{bankman2018always}
D.~Bankman, L.~Yang, B.~Moons, M.~Verhelst, and B.~Murmann, ``An always-on 3.8 microj/86% cifar-10 mixed-signal binary cnn processor with all memory on chip in 28-nm cmos,'' \emph{IEEE Journal of Solid-State Circuits}, vol.~54, no.~1, pp. 158--172, 2018.

\bibitem{moons2018binareye}
B.~Moons, D.~Bankman, L.~Yang, B.~Murmann, and M.~Verhelst, ``Binareye: An always-on energy-accuracy-scalable binary cnn processor with all memory on chip in 28nm cmos,'' in \emph{2018 IEEE Custom Integrated Circuits Conference (CICC)}.\hskip 1em plus 0.5em minus 0.4em\relax IEEE, 2018, pp. 1--4.

\bibitem{hsu2023fully}
Y.-C. Hsu, A.~Kosuge, R.~Sumikawa, K.~Shiba, M.~Hamada, and T.~Kuroda, ``A fully synthesized 13.7 $\mu$j/prediction 88\% accuracy cifar-10 single-chip data-reusing wired-logic processor using non-linear neural network,'' in \emph{Proceedings of the 28th Asia and South Pacific Design Automation Conference}, 2023, pp. 182--183.

\end{thebibliography}

% Biography section
\begin{IEEEbiography}[{\includegraphics[width=1in,height=1.25in,clip,keepaspectratio]{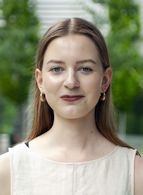}}]{Wiktoria Agata Pawlak}
Wiktoria Agata Pawlak is completing her B.Sc. in Cognitive Science and AI at Tilburg University in the Netherlands, with her focus on neuroscience and machine learning. She holds a student fellowship at the Friedrich Miescher Institute for Biomedical Research (FMI) in Switzerland, contributing to the field of connectomics by developing novel methods for reconstructing neurons to enhance our understanding of brain functions. Additionally, she is a Research Intern at ni2o, a company developing AI-driven brain-computer interfaces for treating neurological disorders and enhancing cognitive abilities. Her work involves improving the biocompatibility of implants, as well as applying machine learning and neuromorphic computing to optimize the performance of the implant's chip integrating it with bioelectrical environment of cells.
\end{IEEEbiography}

\begin{IEEEbiography}[{\includegraphics[width=1in,height=1.25in,clip,keepaspectratio]{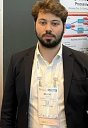}}]{Murat Isik}
Murat Isik received his B.Sc. degree in Electrical and Electronics Engineering from Turkiye in 2021. He subsequently earned his M.Sc. in Electrical and Computer Engineering from Drexel University, Philadelphia, USA, in 2023. Currently, he holds a research scholarship at Stanford University and is pursuing a Ph.D. at Purdue University, focusing on energy estimation projects for various applications. His research interests encompass novel architectures for machine learning acceleration.
\end{IEEEbiography}

\begin{IEEEbiography}[{\includegraphics[width=1in,height=1.25in,clip,keepaspectratio]{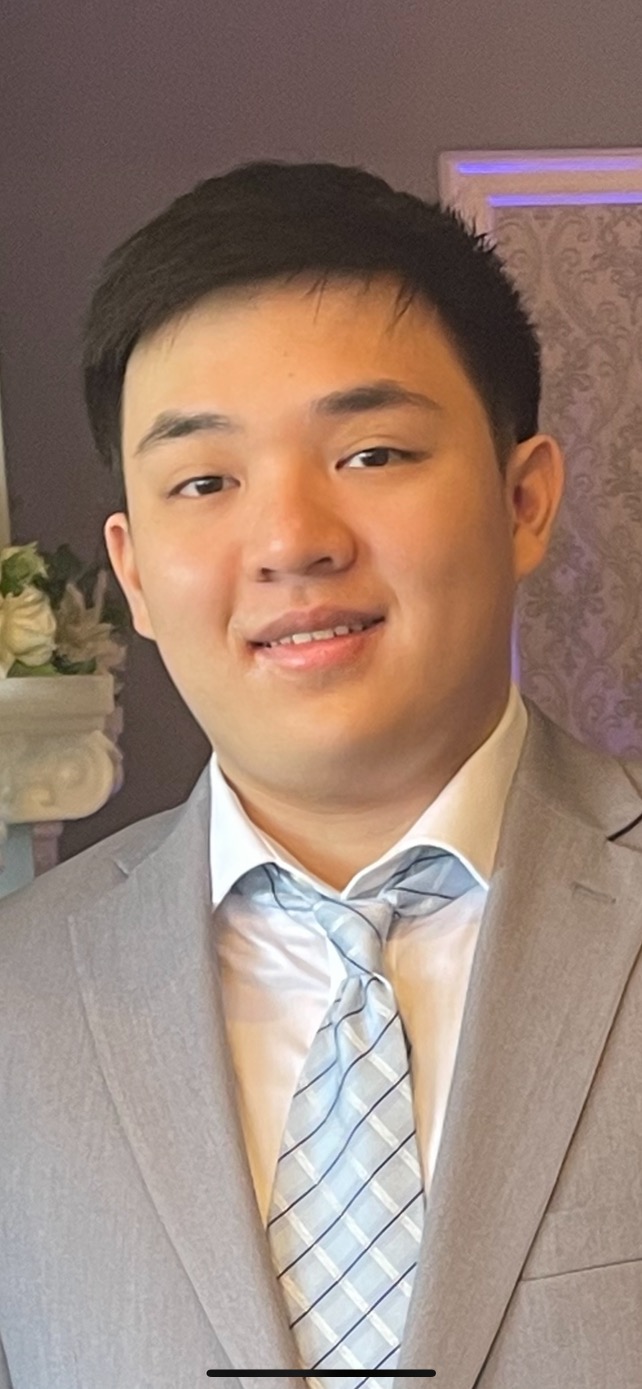}}]{Dexter Le}
Dexter Le is completing his B.Sc. degrees in Computer Science, Electrical, and Computer Engineering from Drexel University, Philadelphia, USA, with a focus on computer networks and high-performance computing. He is a DevOps Engineer for SAP, developing infrastructure as code for enhancing management processes. His work involves designing highly resilient systems to cloud hyper scalers and contributions to the open source community on enhancing operating system security.
\end{IEEEbiography}

\begin{IEEEbiography}[{\includegraphics[width=1in,height=1.25in,clip,keepaspectratio]{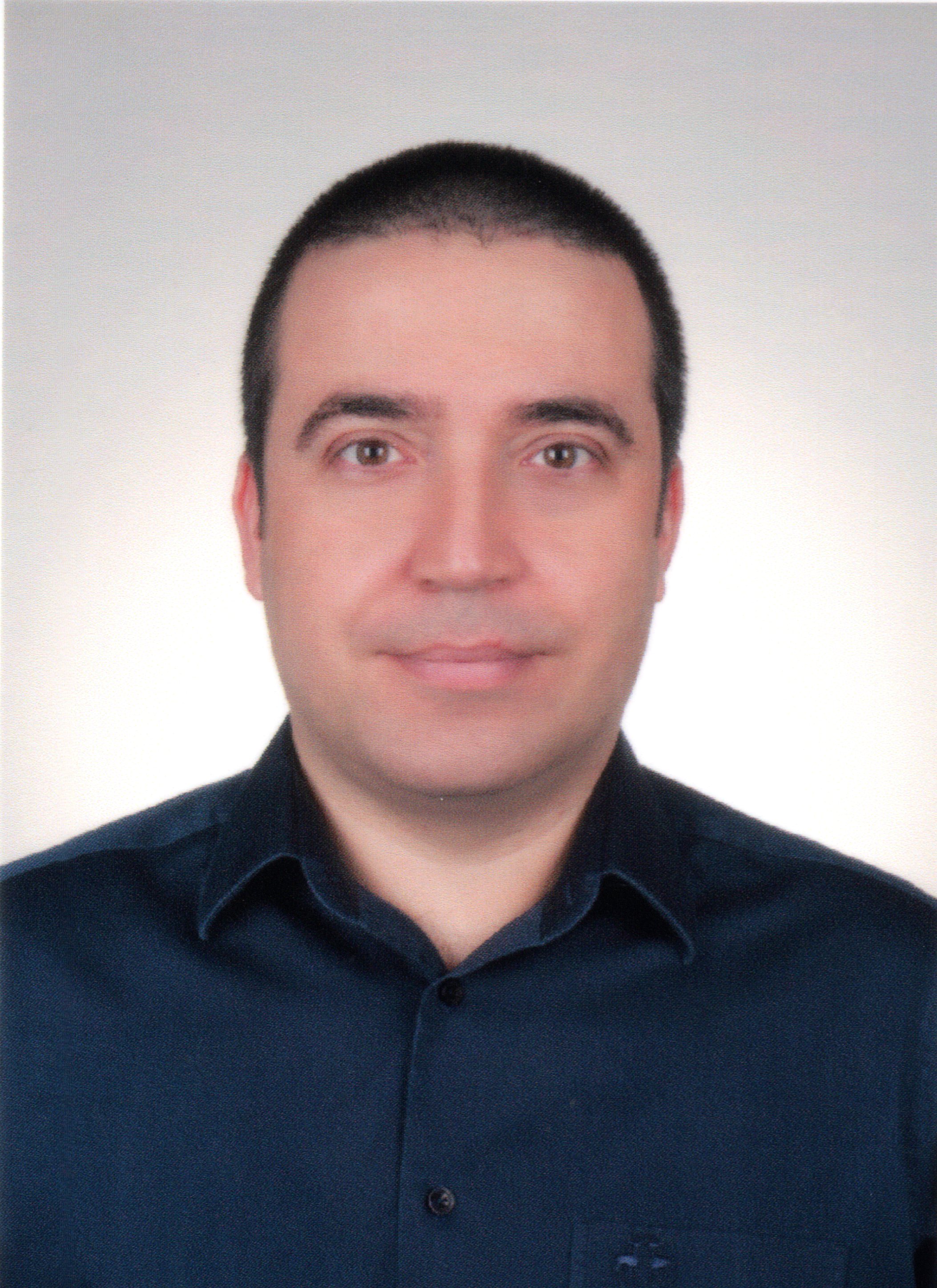}}]{Ismail Can Dikmen}
Ismail Can Dikmen received his B.S. degree in Electronic Engineering from the Turkish Air Force Academy in 2002 and his M.S. degree in Electronic Engineering from the Aeronautics and Space Technologies Institute in 2009. He earned his Ph.D. degree in Electrical and Electronic Engineering from Inonu University in 2022. After serving as a pilot at the Turkish Air Force from 2015 to 2023, he transitioned into various positions, including Lead Engineer at MOTAS Inc. and Head of the Electric and Energy Department at Inonu University. He is currently the Mobility Technologies Lead at the R\&D Center of TEMSA SKODA SABANCI Transportation Vehicles Inc. His research interests span embedded systems development, battery management systems, electric vehicles, machine learning, and neuromorphic hardware integration primarily in automotive. He co-founded SUSTECH Research Development and Consultancy Ltd. in 2018, focusing on battery management systems for electric vehicles. His contributions include patents on electric vehicles battery technologies and IoT-based medical transport devices. His awards and honors include the First Prize in the Most Innovative Scientific-Based Project Category at the RDCONE R\&D and Innovation Awards in 2021.
\end{IEEEbiography}

% End of the document
\end{document}